\documentclass[12pt]{main}

\title{Correlating electron microscopy and synchrotron X-ray measurements: design and implementation of a TEM-holder-compatible vacuum chamber at the Advanced Light Source}
\author[1]{John C. Thomas}
\author[2]{Tara P. Mishra}
\author[2]{Karen C. Bustillo}
\author[1,3]{Alisson S. Thill}
\author[1]{Lucas Kistulentz}
\author[2]{Livia Pugens Matte}
\author[1,4]{Xian'e Li}
\author[1]{Aaron Bostwick}
\author[1]{Martin Keller}
\author[1]{Xin Chen}
\author[1]{Damian Guenzing}
\author[1]{Matthew Roizin-Prior}
\author[3]{Fabiano Bernardi}
\author[1]{Damon English}
\author[1]{Dilworth Y. Parkinson}
\author[1*]{Slavomir Nemsak}
\affil[1]{Advanced Light Source, Lawrence Berkeley National Laboratory, Berkeley, CA 94720, United States of America}
\affil[2]{National Center for Electron Microscopy, Molecular Foundry, Lawrence Berkeley National Laboratory, Berkeley, CA 94720, United States of America}
\affil[3]{Institute of Physics, Federal University of Rio Grande do Sul, Brazil}
\affil[4]{Laboratory of Organic Electronics, Department of Science and Technology (ITN), Linköping University, SE-601 74 Norrköping, Sweden}
\affil[*]{snemsak@lbl.gov}

\usepackage{amsmath} 
\date{}
\begin{document}
\maketitle
\setstcolor{red}
\section*{Abstract}
We present the design of an experimental solution that enables an air-free transfer between a commercial transmission electron microscope (TEM), under high vacuum, and the ambient pressure x-ray photoelectron spectroscopy, x-ray absorption spectroscopy, and x-ray scattering endstation. The setup adds additional capability to APPEXS (ambient-pressure photoemission and x-ray scattering) endstation of beamline 11.0.2.1 at the Advanced Light Source (ALS), where the TEM sample holder can be loaded directly into the ultrahigh vacuum chamber and positioned within the working distance of the hemispherical electron analyzer and x-ray beam. The presented solution is compatible with numerous commercially available sample holders providing different measurement modalities, which can be used for cross-correlative measurements between synchrotron and TEM experiments. Here, we demonstrate the use of this setup by synchrotron spectroscopic measurements in both UHV conditions and higher pressures directly compared with imaging of the same sample in the TEM without exposure to air.

\section*{Introduction}
Understanding the structure, morphology, and chemical state of a functional material under realistic operating conditions is a central goal of modern catalysis, energy, and environmental science. For heterogeneous catalysts in particular, the composition, oxidation state, and morphology of the active surface can differ substantially between the idealized ultrahigh vacuum (UHV) conditions of classical surface science and the elevated pressures and temperatures of an operating reactor and between well-defined single-crystal model systems to more complex nonuniform materials~\cite{Salmeron_2008, Starr_2013}. Bridging these gaps requires analytical tools that can interrogate a sample \emph{in situ}, ideally correlating complementary observables on the same material under the same conditions.

X-ray photoelectron spectroscopy (XPS) is uniquely suited for probing the surface chemistry of such systems because it simultaneously reports on the elemental composition and surface chemical state. Conventional XPS is nonetheless restricted to vacuum by the short inelastic mean free path of photoelectrons in gases. In the near-ambient regime, however, ambient pressure photoelectron spectroscopy is capable of probing the sample surface state in a gas environment and under reaction conditions~\cite{Bluhm_2007}. The decisive advance for modern instruments was the addition of a differentially pumped electrostatic lens system, which collects and refocuses photoelectrons that pass through a small aperture close to the sample, dramatically increasing the collected signal at millibar pressures~\cite{Ogletree_2002, Ogletree_2009}. This method, called ambient pressure XPS (AP-XPS), has since been applied to problems ranging from catalytic oxidation reactions~\cite{Bluhm_2004, Toyoshima_2015} to electrochemical interfaces~\cite{Crumlin_2013}, and reaction-driven restructuring of bimetallic nanoparticles has been observed directly by these methods~\cite{Tao_2008}.

While AP-XPS provides surface chemical information, it does not resolve the real-space, nanoscale morphology of individual particles. Transmission electron microscopy (TEM), and its \emph{in situ} and environmental variants, supplies precisely this complementary structural information, and a range of environmental cells and micro-electromechanical-system (MEMS) sample holders have been developed to expose specimens to gases and liquids inside the microscope~\cite{Su_2023}. The most direct way to correlate chemistry with structure is therefore to measure the same sample by both techniques. Recent work has pursued this in several ways: by designing electron-transparent microreactors that permit correlative XPS and TEM on the same type of sample under matched conditions~\cite{10.1063/5.0308362}, by transferring a single \emph{in situ} cell between a hard x-ray nanoprobe beamline and an electron microscope without changing the gas environment~\cite{Cheng_2022}, and by employing ultrathin two-dimensional membranes that isolate a reactive environment from the analyzer while remaining transparent to both electrons and photoelectrons~\cite{Weatherup_2018}. A common thread is the recognition that using shared cell geometries and commercially available, TEM-compatible sample holders makes the two techniques far more readily comparable.

At the ALS, the coupling of a commercial TEM sample manipulator, a goniometer, to a synchrotron endstation was previously demonstrated for soft x-ray ptychography, where the Thermo Fisher compatible manipulator provided the nanometer-scale positioning required for coherent diffractive imaging and opened the door to cross-correlative measurements using both soft x-rays and electrons on identical specimens~\cite{Celestre_2017}. Building on that concept, we describe here the design and implementation of a UHV chamber for APPEXS endstation\cite{10.1063/5.0044162} at ALS beamline 11.0.2.1 that accepts the same class of commercial TEM holders. The chamber enables a mounted sample to be introduced into the analysis position, aligned to within 500 $\mu$m of the differentially pumped electrostatic lens cone, measurements under UHV and elevated pressure conditions, and then removed for structural imaging of the very same region in a TEM at the neighboring National Center for Electron Microscopy (NCEM) within the Molecular Foundry. This workflow provides both 1) rapid sample exchange and 2) a direct, same-sample bridge between the chemical-state sensitivity of AP-XPS and the structural resolution of TEM.

\section*{Mechanical Design}
The central design requirement was to reconcile two normally incompatible geometries: the compact, side-entry stage of a commercial TEM holder, and the aperture-limited collection geometry of a differentially pumped AP-XPS analyzer, in which useful signal is obtained only when the sample sits within a few hundred micrometers of the first lens aperture~\cite{Ogletree_2002, Grass_2010}. The chamber was therefore designed around three coupled subsystems: (i) a Thermo Fisher Compustage sample transfer path; (ii) a welded mating sleeve that mounts the Compustage (preserving location and vacuum requirements); and (iii) an air-free sample holder that can be used between AP-XPS and TEM with minimal exposure to atmosphere (Fig.~\ref{fig1:design}).

The analysis chamber is an ultrahigh vacuum vessel that is connected to the differentially pumped electrostatic lens of the hemispherical analyzer. The sample is brought to the focal point of the lens, where the entrance aperture (nominally sub-millimeter in diameter) defines both the pressure step between the sample region and the first pumping stage and the acceptance of the lens~\cite{Ogletree_2002}. The tight, sub-500 $\mu$m sample-to-cone working distance shown in Fig.~\ref{fig2:APXPS} is the single most demanding tolerance in the assembly: it sets the allowable play in the manipulation stage, allowable sample geometry, and the alignment budget for the incident x-ray beam, which must illuminate the sample within the small volume imaged by the lens. To achieve this while retaining the ability to insert and retract a commercial holder, the stage provides coarse translation for holder exchange together with fine piezo positioners, so that the sample can be walked up to the cone without risk of contact. The four Compustage axes (X, Y, Z, and $\alpha$ tilt) are servo-driven and controlled by a Galil DMC-4143 four-axis motion controller with an integrated AMP-43140 linear amplifier (Galil Motion Control). The controller is integrated into the ALS Beamline Control System, which provides motor control, scanning, and data acquisition consistent with other ALS endstations.

Because commercial TEM holders are designed to be inserted into the goniometer of a microscope rather than into a photoemission endstation, the transfer mechanics were adapted to the holder rather than the reverse. This preserves the key advantage of the approach, where standard off-the-shelf holders (including environmental and \emph{in situ} variants) can be used without modification and exactly as envisaged for the earlier TEM-manipulator-based ptychography instrument at the ALS~\cite{Celestre_2017}. It also ensures that a sample never has to be dismounted from its holder between the spectroscopy and microscopy measurements. Maintaining the sample on a single holder throughout is what makes true same-region and same-history correlation possible and is essential for air-sensitive specimens, for which any transfer step is an opportunity for contamination or oxidation.

Design details of the chamber and the holder interface are shown schematically in Fig.~\ref{fig1:design}, and the sample-to-cone geometry in the loaded and unloaded configurations is shown in Fig.~\ref{fig2:APXPS}. The dimensioned sleeve weldment itself is shown in Supplementary Figures 1 and 2, where the distance of the sleeve port axis to cone, orientation, and height are critical during fabrication.

\section*{Results}
To demonstrate the combined workflow, a supported bimetallic Pd-Pt nanofoam, which is used for hydrogen storage \cite{10.1039/d5ta04270d}, was mounted on a commercial FEI-compatible air-free TEM holder (manufactured by ZoNexus LLC), introduced into the chamber, and positioned at the analyzer cone as shown in Fig.~\ref{fig2:APXPS}. With the sample retracted, the cone geometry is clearly visible; with the sample extended, the sub-500 $\mu$m working distance between the sample surface and the cone is achieved, placing the illuminated region within the acceptance of the differentially pumped lens.

Core-level spectra were acquired for the Pd $3d$ and Pt $4f$ regions before and after exposure to hydrogen (Fig.~\ref{fig3:Spectra}). The two spin-orbit-split doublets characteristic of each element, Pd $3d_{5/2}$/$3d_{3/2}$ and Pt $4f_{7/2}$/$4f_{5/2}$, are resolved in both conditions. Upon H$_{\rm 2}$ exposure, both Pd and Pt spectra suggest the reduction of the materials, where the Pd$^{0}$ and Pt$^{0}$ peaks are now predominant, and both the Pd$^{2+}$ and Pt$^{2+}$ intensities have diminished. For Pd, reduction to Pd$^{0}$ was nearly complete. In contrast, Pt exhibits both Pt$^{2+}$, though lower intensity, and Pt$^{0}$. This element-specific behavior is expected for Pd–Pt systems, where differences in hydrogen uptake (dominantly in Pd), relative oxidation/reduction tendencies, and electronic interaction between the two metals are known to drive differential chemical-state shifts and reaction-induced restructuring~\cite{Tao_2008, Kobayashi_2010}. In fact, the previous study reported very similar behavior of the Pd and Pt components before and after H$_{\rm 2}$ exposure\cite{10.1039/d5ta04270d}. Moreover, evidence was obtained indicating that H$_{\rm 2}$ exposure induces changes in the atomic arrangement of the nanostructure. The O $1s$ region shows an attenuation of the region associated with the nanofoam oxide under 1 Torr H$_{\rm 2}$. This further confirms the reduction of the oxide phase. 

Following the spectroscopic measurements, the sample was transferred, using the same holder, to a TEM at NCEM, and the resulting image is shown in Fig.~\ref{fig4:TEM}. Here, both parallel beam TEM and annular dark-field scanning transmission electron microscopy (ADF-STEM) images were collected. The lowest magnification image showing the full holey silicon nitride membrane window, is shown in Fig.~\ref{fig4:TEM} (a). These windows enable correlative AP-XPS and TEM on the same nanofoam sample because the thin, electron-transparent graphene coating of SiN is robust while remaining suitable for both conventional TEM and ADF-STEM imaging, allowing bright-field morphology and Z-contrast elemental distribution (Pd/Pt vs. lighter oxide/support phases) to be directly compared with the chemical-state information from XPS. As the specimen remained on its holder and was transferred without exposure to air, the structure imaged in the microscope corresponds to the same material interrogated spectroscopically at the beamline. This provides a direct, same-sample correlation between the surface chemical state measured by AP-XPS and the nanoscale morphology resolved by TEM.

\section*{Discussion}
The results establish a practical route to same-sample, cross-correlative AP-XPS and TEM using commercial, unmodified sample holders. The principal advantage of the approach is that it removes the need to fabricate a bespoke, dual-purpose cell or to transfer the specimen between incompatible mounts: the identical holder that is measured spectroscopically is subsequently imaged in the microscope. This directly addresses the correlation problem that motivates much recent instrument development, where the central difficulty is ensuring that the two measurements probe the same particles under comparable conditions and with identical synthesis and treatment history~\cite{Cheng_2022, 10.1063/5.0308362}. By adopting the Thermo Fisher compatible holder standard already used for TEM-manipulator-based x-ray microscopy at the ALS~\cite{Celestre_2017}, the present chamber leverages an existing ecosystem of environmental and \emph{in situ} holders and makes the two techniques immediately comparable in geometry and sample format~\cite{Weatherup_2018}.

The Pd-Pt measurements illustrate the scientific value of the combination. Chemical-state changes observed by AP-XPS, such as the unique Pd and Pt responses to hydrogen, are difficult to interpret without knowledge of whether the underlying particles have also restructured, segregated, or sintered, phenomena that are well documented for bimetallic nanoparticles under reactive gases~\cite{Tao_2008, Kobayashi_2010}. Conversely, morphological details seen in the microscope are most informative when paired with the surface chemistry that accompanies them. Same-sample correlation ties the two together. The air-free transfer step is essential to this: without it, exposure to the atmosphere between measurements would alter the surface chemical state and could obscure the very changes of interest.

Several extensions follow naturally from the present design. The chamber is compatible with the range of commercial holders that provide heating, biasing, and gas or liquid environments, opening the possibility of closer matching of conditions between the spectroscopy and microscopy measurements. More broadly, the approach is applicable well beyond catalysis to any air-sensitive or environmentally responsive system for which correlating surface chemistry with nanoscale structure on an identical specimen is decisive, such as with battery and electrochemical interfaces, environmental surfaces, and energy materials. Ultimately, the mating design is compatible with most standard UHV setups and can be integrated during upgrades of several other endstations at the ALS and other synchrotron facilities.

\section*{Materials and Methods}
\subsection*{Sample Preparation}
The Pd-Pt nanofoam sample was prepared with nominal compositions of Pt$_{50}$Pd$_{50}$ following the synthesis method described in a previous work\cite{10.1039/d5ta04270d}. In order to obtain the nanofoam structure the monoolein/water system was used as a template. The Pd and Pt were obtained from chemical reduction of potassium hexachloroplatinate (K$_{2}$PtCl$_{6}$, 99.9\%) and palladium acetate (Pd(OCOCH$_{3}$)$_{2}$, 99.9\%), using glucose (C$_{6}$H$_{12}$O$_{6}$, 99.9\%) as the reducing agent. For the template, a technical-grade monoolein, containing approximately 40\% 1-oleoyl-rac-glycerol, with diolein and triolein as the main impurities, was used. First, K$_{2}$PtCl$_{6}$ was dissolved together with 38.9 mg of glucose in 7.2 mL of Milli-Q water. Separately, the required amount of Pd(OCOCH$_{3}$)$_{2}$ was incorporated into 20.0 g of monoolein. This mixture was placed in an ultrasonic bath for 5 min while being continuously homogenized with a spatula. This mixture was subsequently added to the aqueous Pt salt solution. The resulting system was maintained at 80 $^\circ$C in a water bath for 1 h and manually homogenized at 5 min intervals using a spatula. During this step, the mixture progressively darkened and ultimately became black, indicating the formation of the reduced metallic phase. After that, the mixture was allowed to rest at room temperature for 2 h. Then, it was transferred to a crucible for the calcination procedure at 500 $^\circ$C for 4 h. After calcination, a dark powder was obtained. As a final step, the powder was washed 10 times with Milli-Q water, with each washing step involving centrifugation at 3000 rpm for 10 min. The final material was dried under vacuum in a desiccator for 24 h.

\subsection*{Ambient pressure x-ray photoelectron spectroscopy}
AP-XPS measurements were performed at APPEXS endstation of beamline 11.0.2.1 of the Advanced Light Source (ALS), Lawrence Berkeley National Laboratory\cite{10.1063/5.0044162}. The endstation employs a SPECS Phoibos near-ambient pressure hemispherical electron analyzer coupled to a differentially pumped electrostatic lens system, following the design principles established for photoemission in the millibar range, in which photoelectrons pass through a sub-millimeter aperture close to the sample and are collected and refocused by successive differentially pumped lens stages held at progressively lower pressure~\cite{Ogletree_2002, Ogletree_2009, Grass_2010}. The sample, mounted on a commercial TEM holder, was positioned within working distance of the hemispherical analyzer (Fig.~\ref{fig2:APXPS}). Core-level spectra of the Pd $3d$ and Pt $4f$ regions were recorded under ultrahigh vacuum and during introduction of 1 Torr H$_{\rm 2}$ into the analysis chamber. Binding energies and the assignment of chemical states were referenced to standard values for metallic and oxidized Pd and Pt. Additional details can be found in prior work\cite{10.1039/d5ta04270d}.

\subsection*{Transmission electron microscopy}
Measurements were performed at the National Center for Electron Microscopy (NCEM) at the Molecular Foundry, Lawrence Berkeley National Laboratory. Following the AP-XPS measurements in H$_{\rm 2}$ atmosphere, the sample was transferred to the microscope on the same commercial holder used at the beamline, without exposure to air, so that the region imaged corresponds to the material characterized spectroscopically (Fig.~\ref{fig4:TEM}). The holder is compatible with most Thermo Fisher TEMs and in this case it was inserted into a ThemIS operating at 300kV. Figs. ~\ref{fig4:TEM} (a and b) show parallel beam TEM images at low magnification. In Figs. ~\ref{fig4:TEM} (c and d), the particles were imaged in microprobe STEM mode with a semi-convergence angle of 0.5 mrad at low magnification with the objective lens and C3 lens turned off. An annular dark-field (ADF) detector acquired the scattered electrons; the contrast is both from Bragg scattered electrons and from interactions sensitive to mass/thickness. As can be seen from Figs. ~\ref{fig4:TEM} (c and d) where the grid holes appear to be ellipses instead of circles, there is astigmatism in the low-magnification STEM illumination mode.

\subsection*{Data availability} 
Data that support the findings of this study are available from the corresponding authors on reasonable request.


\begin{thebibliography}{10}

\bibitem{Salmeron_2008}
Salmeron, M. \& Schl{\"o}gl, R. Ambient pressure photoelectron spectroscopy:
  A new tool for surface science and nanotechnology. {\em Surf. Sci.
  Rep.} \textbf{63}, 169 (2008).

\bibitem{Starr_2013}
Starr, D.E., Liu, Z., H{\"a}vecker, M., Knop-Gericke, A. \& Bluhm, H.
  Investigation of solid/vapor interfaces using ambient pressure x-ray
  photoelectron spectroscopy. {\em Chem. Soc. Rev.} \textbf{42}, 5833 (2013).

\bibitem{Bluhm_2007}
Bluhm, H., H{\"a}vecker, M., Knop-Gericke, A., Kiskinova, M., Schl{\"o}gl, R. \&
  Salmeron, M. In situ x-ray photoelectron spectroscopy studies of gas/solid
  interfaces at near-ambient conditions. {\em MRS Bull.} \textbf{32}, 1022 (2007).

\bibitem{Ogletree_2002}
Ogletree, D.F., Bluhm, H., Lebedev, G., Fadley, C.S., Hussain, Z. \&
  Salmeron, M. A differentially pumped electrostatic lens system for
  photoemission studies in the millibar range. {\em Rev. Sci.
  Instrum.} \textbf{73}, 3872 (2002).

\bibitem{Ogletree_2009}
Ogletree, D.F., Bluhm, H., Hebenstreit, E.D. \& Salmeron, M. Photoelectron
  spectroscopy under ambient pressure and temperature conditions. {\em
  Nucl. Instrum. Methods Phys. Res. A: Accel. Spectrom. Detect. Assoc. Equip.} \textbf{601}, 151 (2009).

\bibitem{Bluhm_2004}
Bluhm, H., H{\"a}vecker, M., Knop-Gericke, A., Kleimenov, E., Schl{\"o}gl, R., Teschner, 
  D., Bukhtiyarov, V.I., Ogletree, D.F. \& Salmeron, M. Methanol oxidation on a copper 
  catalyst investigated using in situ x-ray photoelectron
  spectroscopy. {\em J. Phys. Chem. B} \textbf{108}, 14340 (2004).

\bibitem{Toyoshima_2015}
Toyoshima, R. \& Kondoh, H. In situ observations of catalytic surface
  reactions with soft x-rays under working conditions. {\em J. Phys.: Condens. Matter} \textbf{27}, 083003 (2015).

\bibitem{Crumlin_2013}
Crumlin, E.J., Bluhm, H. \& Liu, Z. In situ investigation of electrochemical
  devices using ambient pressure photoelectron spectroscopy. {\em J. Electron Spectrosc. Relat. Phenom.} \textbf{190}, 84 (2013).

\bibitem{Tao_2008}
Tao, F., Grass, M.E., Zhang, Y., Butcher, D.R., Renzas, J.R., Liu, Z.,
  Chung, J.Y., Mun, B.S., Salmeron, M. \& Somorjai, G.A. Reaction-driven
  restructuring of Rh-Pd and Pt-Pd core-shell nanoparticles. {\em Science} \textbf{322}, 932 (2008).

\bibitem{Su_2023}
Qu, J., Sui, M. \& Li, R. Recent advances in in-situ transmission electron
  microscopy techniques for heterogeneous catalysis. {\em iScience} \textbf{26}, 107072 (2023).

\bibitem{10.1063/5.0308362}
Cai, J., Yao, Z., Zang, Y., He, Y., Gu, Y., Liu, K., Liu, Z. \& Wang, Z.-J.
  Correlative in situ x-ray photoelectron spectroscopy and transmission
  electron microscopy characterization under identical reaction conditions.
  {\em Rev. Sci. Instrum.} \textbf{97}, 055202 (2026).

\bibitem{Cheng_2022}
Parker, J.E., Gomez-Gonzalez, M., Lishout, Y.V., Islam, H., Martin, D.D., Ozkaya, D., 
  Quinn, P.D. \& Schuster, M.E. A cell design for correlative hard x-ray
  nanoprobe and electron microscopy studies of catalysts under in situ
  conditions. {\em J. Synchrotron Radiat.} \textbf{29} 479 (2022).

\bibitem{Weatherup_2018}
Weatherup, R.S. 2D material membranes for operando atmospheric pressure
  photoelectron spectroscopy. {\em Top. Catal.} \textbf{61}, 2085 (2018).

\bibitem{Celestre_2017}
Celestre, R., Nowrouzi, K., Shapiro, D.A., Denes, P., Joseph, J.M., Schmid, A. \&
  Padmore, H.A. Nanosurveyor 2: A compact instrument for nano-tomography at
  the Advanced Light Source. {\em J. Phys. Conf. Ser.} \textbf{849}, 012047 (2017).

\bibitem{10.1063/5.0044162}
Kersell, H., Chen, P., Martins, H., Lu, Q., Brausse, F., Liu, B.-H., Blum, M., Roy, S.,
  Rude, B., Kilcoyne, A., Bluhm, H. \& Nemšák, S. Simultaneous ambient
  pressure x-ray photoelectron spectroscopy and grazing incidence x-ray
  scattering in gas environments. {\em Rev. Sci. Instrum.} \textbf{92}, 044102 (2021).

\bibitem{Grass_2010}
Grass, M.E., Karlsson, P.G., Aksoy, F., Lundqvist, M., Wannberg, B., Mun, B.S.,
  Hussain, Z. \& Liu, Z. New ambient pressure photoemission endstation at
  Advanced Light Source beamline 9.3.2. {\em Rev. Sci. Instrum.} \textbf{81}, 053106 (2010).

\bibitem{10.1039/d5ta04270d}
Khan, W., Thill, A.S., Girotto, G.Z., Vogt, M.A.H., Escudero, C., Poletto, F.,
  Perez-Dieste, V. \& Bernardi, F. Nanoarchitectured Pt–Pd foams as novel
  hydrogen reservoirs through Pt–H bonding. {\em J. Mater. Chem. A} \textbf{14}, 962 (2026).

\bibitem{Kobayashi_2010}
Kobayashi, H., Yamauchi, M., Kitagawa, H., Kubota, Y., Kato, K. \& Takata, M.
  Atomic-level Pd-Pt alloying and largely enhanced hydrogen-storage capacity
  in bimetallic nanoparticles reconstructed from core/shell structure by a
  process of hydrogen absorption/desorption. {\em J. Am. Chem. Soc.}\textbf{132}, 5576 (2010).

\end{thebibliography}

\subsection*{Acknowledgements}
The authors thank Hendrik Ohldag, Andrew Doran, and Kenneth A. Goldberg (Advanced Light Source) for helpful discussions during instrumentation development, Alpesh Shukla (ZoNexus LLC) for guidance on high-throughput and air-free TEM holders, and Rohan Dhall and Peter Ercius (Molecular Foundry) for sample mounting advice. X.L. acknowledges support from an ALS Collaborative Postdoctoral Fellowship and the Swedish Research Council (grant no. 2023-00357). F.B. thanks CNPq for research grant support. Work was performed at the Advanced Light Source and the Molecular Foundry and supported by the Office of Science, Office of Basic Energy Sciences, of the U.S. Department of Energy under contract no. DE-AC02-05CH11231.

\subsection*{Author Contributions}
J.C.T., T.P.M., L.K., M.R.-P., D.Y.P., A.B., and S.N. were part of the instrument design team. J.C.T., A.S.T., X.L., and S.N. carried out AP-XPS measurements. A.S.T. and K.C.B. carried out TEM measurements. M.K., X.C., D.G., and D.E. supported the integration and commissioning of the motor stage. All authors discussed the results and contributed toward the manuscript.
  
\subsection*{Competing interests} The authors declare
 that they have no competing interests. 

\renewcommand{\figurename}{Fig.}
\renewcommand{\thefigure}{1}
\begin{figure}[H]
    \centering
    \includegraphics[width = 1 \linewidth]{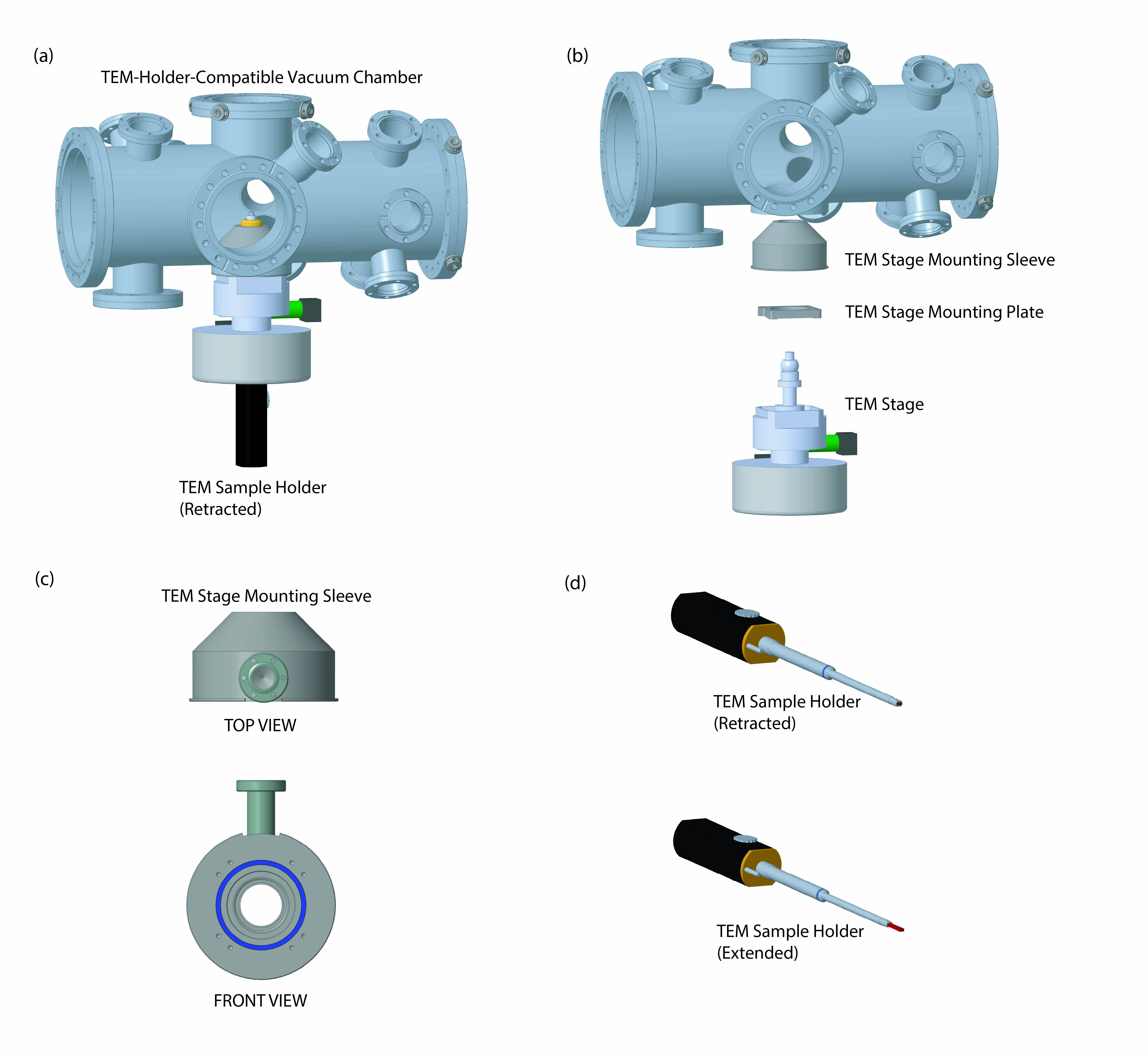}
    \caption{\textbf{Customized x-ray photoelectron spectroscopy chamber for TEM sample holder mounting.} (a) The assembled version of the chamber where a TEM sample holder can be inserted into a Thermo Fisher Compustage. (b) Exploded assembly showing mating components for the stage onto a TEM stage mounting sleeve, which is fabricated separately and then welded to the main chamber. (c) Top and front views of the TEM mounting sleeve. A pumping port (DN16CF) is welded to the sleeve to reach needed vacuum level (<$1 \times 10^{-3}$ Torr) on the Compustage during sample transfer. TEM stage mounting plate tapped holes are mirrored along two planes, such that the Compustage can be mounted in four different configurations. An o-ring groove (blue) is included so a plate, with sufficient surface finish, can be mounted to the sleeve to reach UHV vacuum pressure with the Compustage unmounted. This enables additional ports of access to measure samples held in other configurations (not shown). (d) A high-throughput TEM holder, with double tilt on two of the sample locations, is employed to enable air-free transfer of samples between AP-XPS and TEM tool sets. 
     }
    \label{fig1:design}
\end{figure}

\renewcommand{\thefigure}{2}
\begin{figure}[H]
    \centering
    \includegraphics[width = 1 \linewidth]{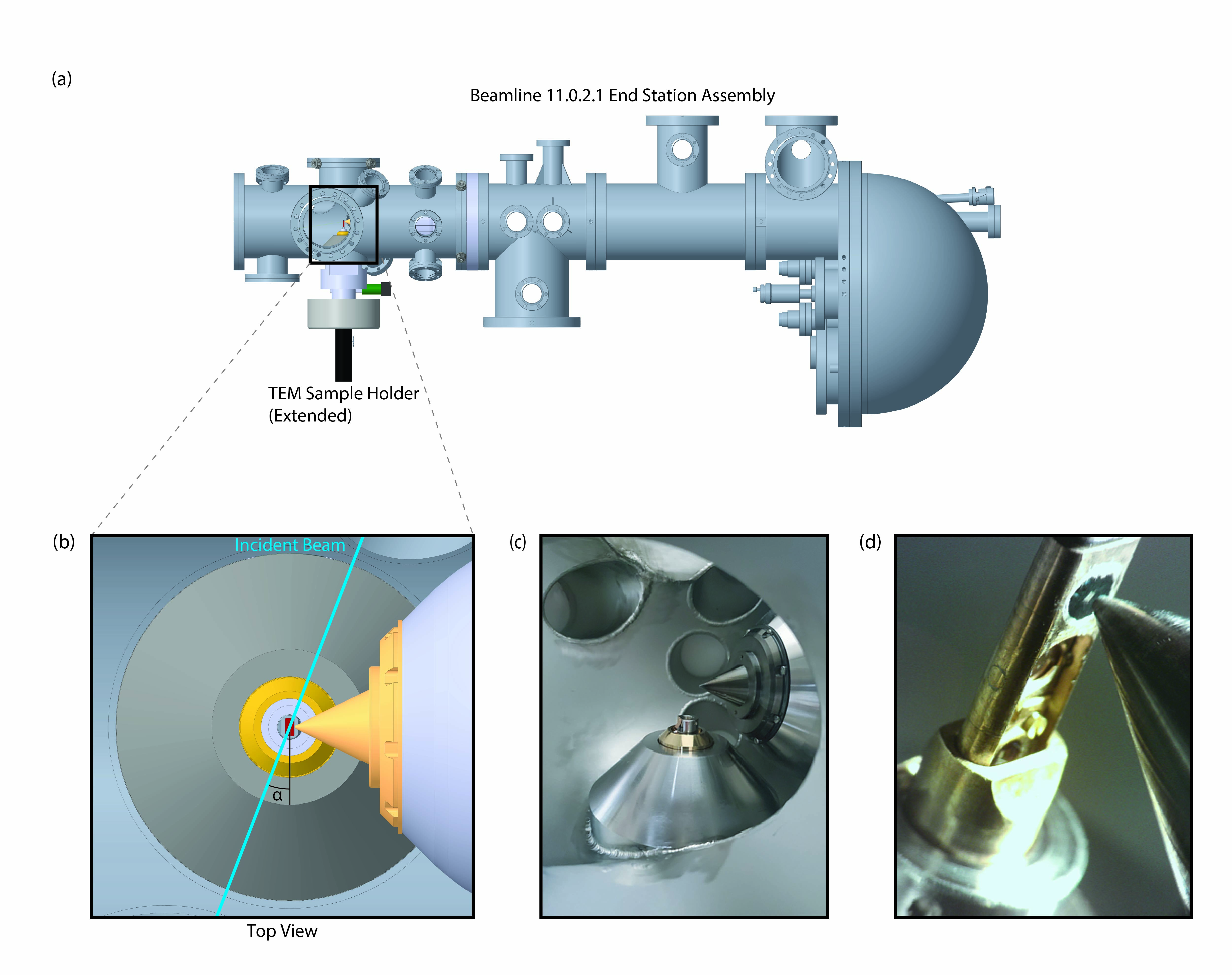}
     \caption{\textbf{Endstation assembly of TEM-compatible vacuum chamber.} (a) Photoelectron collection is performed with a hemispherical electron analyzer coupled to the sample region through a differentially pumped electrostatic lens system. (b) The entrance to the analyzer is a tapered aperture cone positioned within approximately 500 $\mu$m of the sample surface. X-rays are incident upon the sample at $\sim$20$^\circ$ grazing incidence, and the resulting photoelectrons are collected by the hemispherical analyzer (near normal to the sample surface). (c) Image of a mounted high-throughput holder that is retracted. (d) The holder is also shown extended, where the Compustage brings the sample within the required distance to the aperture cone for measurement.}
    \label{fig2:APXPS}
\end{figure}

\renewcommand{\thefigure}{3}
\begin{figure}[H]
    \centering
    \includegraphics[width = 1 \linewidth]{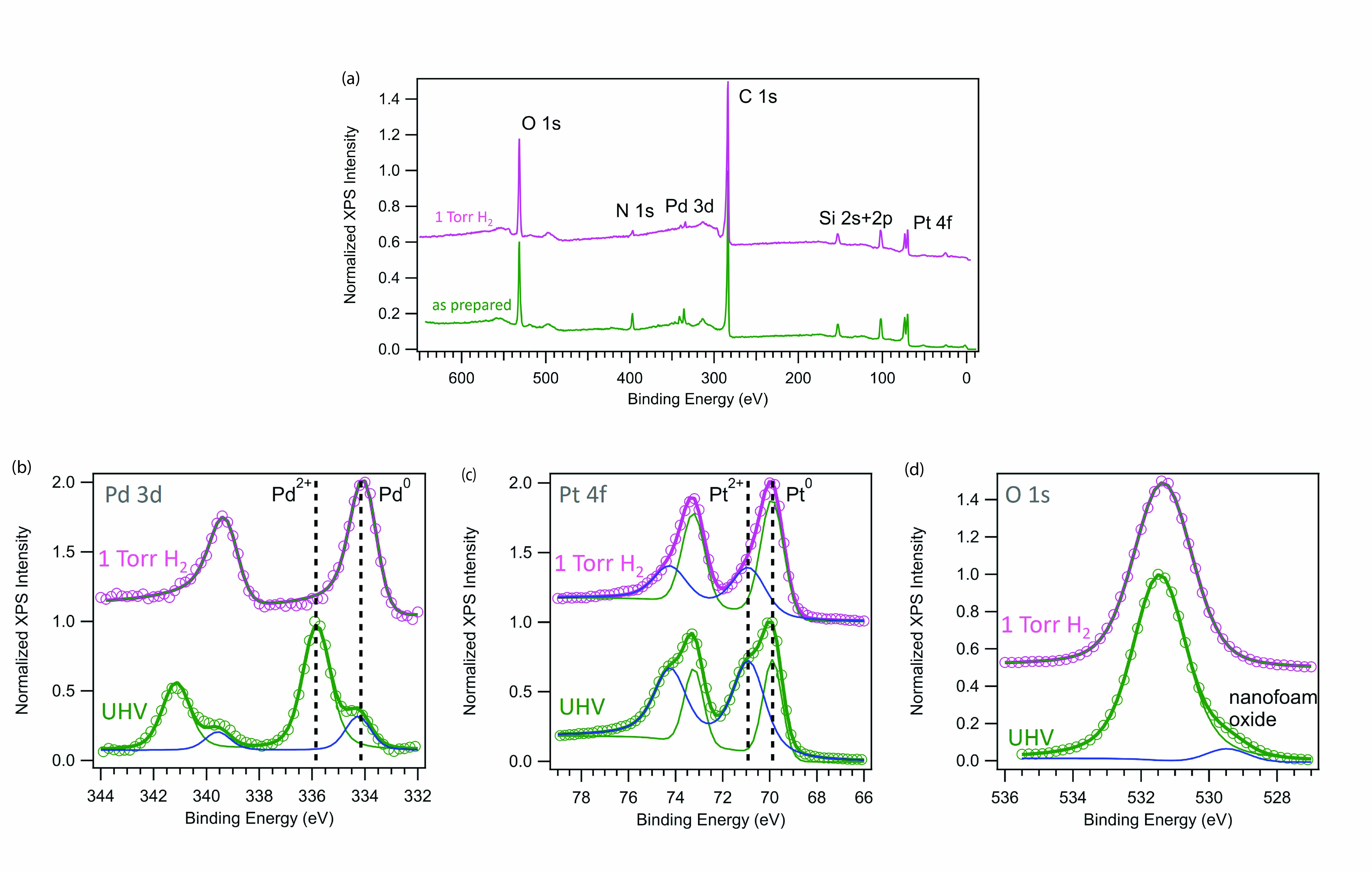}
    \caption{\textbf{X-ray photoelectron spectroscopy}. (a) Survey XPS spectra acquired on the Pt$_{50}$Pd$_{50}$ nanofoam sample. The high resolution measurements show the (b) Pd $3d$, (c) Pt $4f$ and (d) O $1s$ XPS spectra. All measurements are obtained before (green) and during (magenta) H$_{\rm 2}$ exposure.}
    \label{fig3:Spectra}
\end{figure}

\renewcommand{\thefigure}{4}
\begin{figure}[H]
    \centering
    \includegraphics[width = 1 \linewidth]{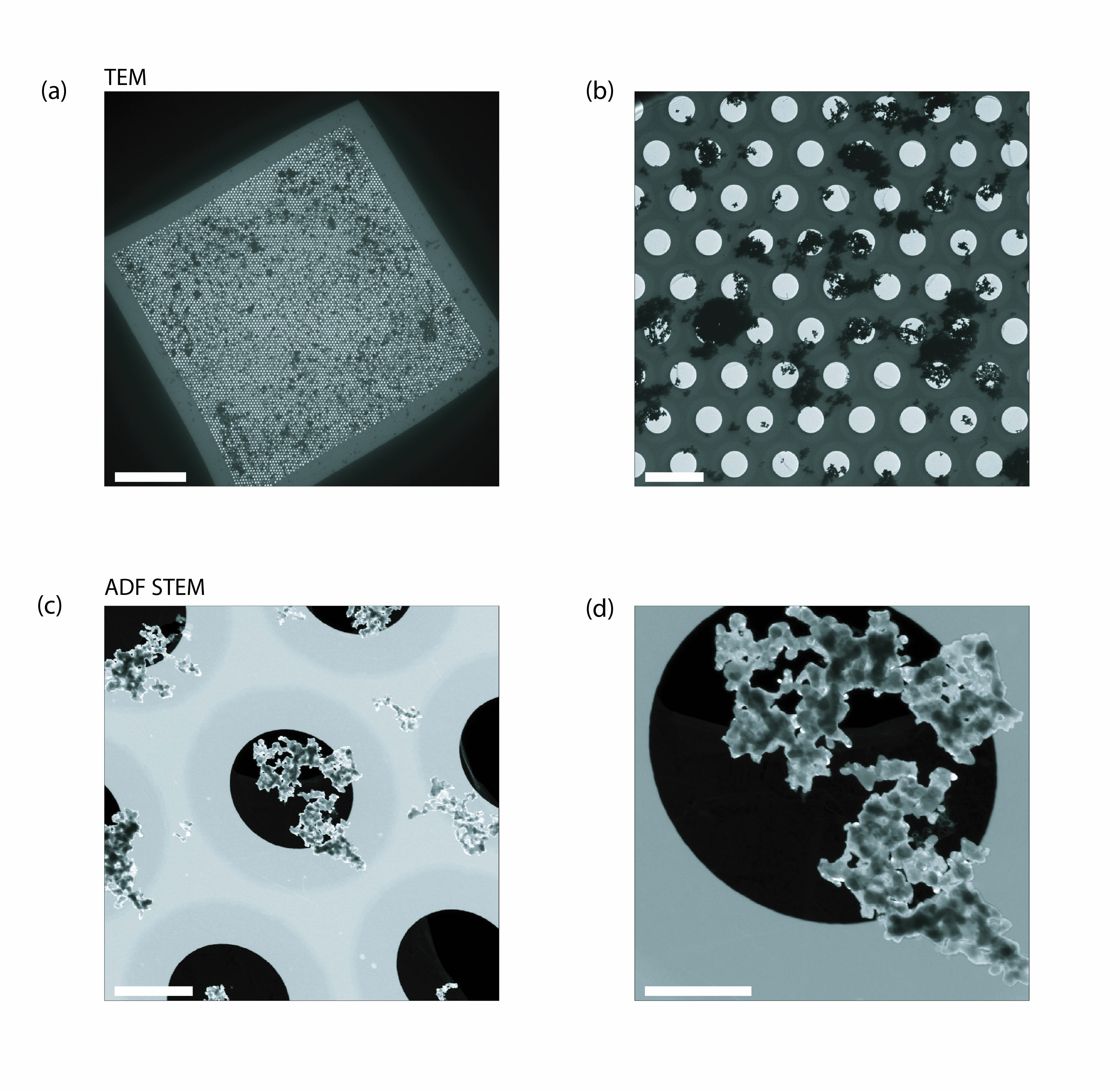}
    \caption{\textbf{Transmission electron microscopy.} Data shown were collected after air free transfer of the sample from the beamline to the TEM. (a) Low-magnification parallel illumination TEM image showing particle coverage on one grid square. Scale bar, 100 $\mu$m. (b) Low-magnification parallel illumination TEM image showing periodic array of 2 $\mu$m holes in TEM support. Scale bar, 5 $\mu$m. (c) and (d) Annular dark-field low-magnification scanning-TEM (STEM) images showing particle morphology. Scale bars, 1 $\mu$m and 500 nm, respectively.}
    \label{fig4:TEM}
\end{figure}

\end{document}


\maketitle

\section*{Supplementary Figures}
\renewcommand{\figurename}{Supplementary Fig.}
\renewcommand{\thefigure}{1}
\begin{figure}[H]
    \centering
    \includegraphics[width = 1 \linewidth]{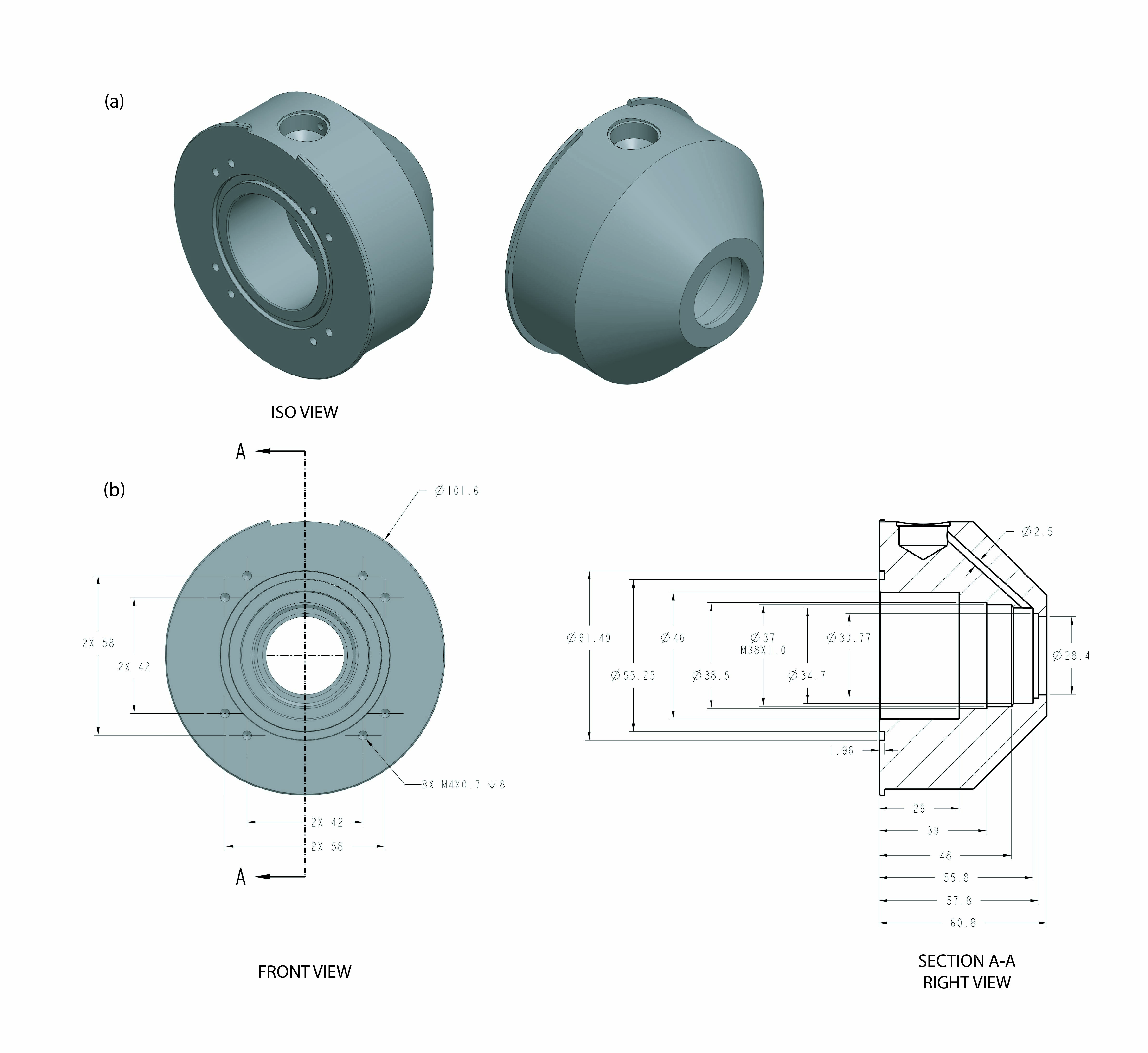}
     \caption{\textbf{Dimensioned TEM stage mounting sleeve.} (a) Isometric views show the front and back of the sleeve before weldment. (b) Front and Right views indicate sleeve dimensions. Here, section A-A shows the concentric stepped bore that is critical to mount the stage.}
    \label{figS1:GDT}
\end{figure}

\renewcommand{\thefigure}{2}
\begin{figure}[H]
    \centering
    \includegraphics[width = 1 \linewidth]{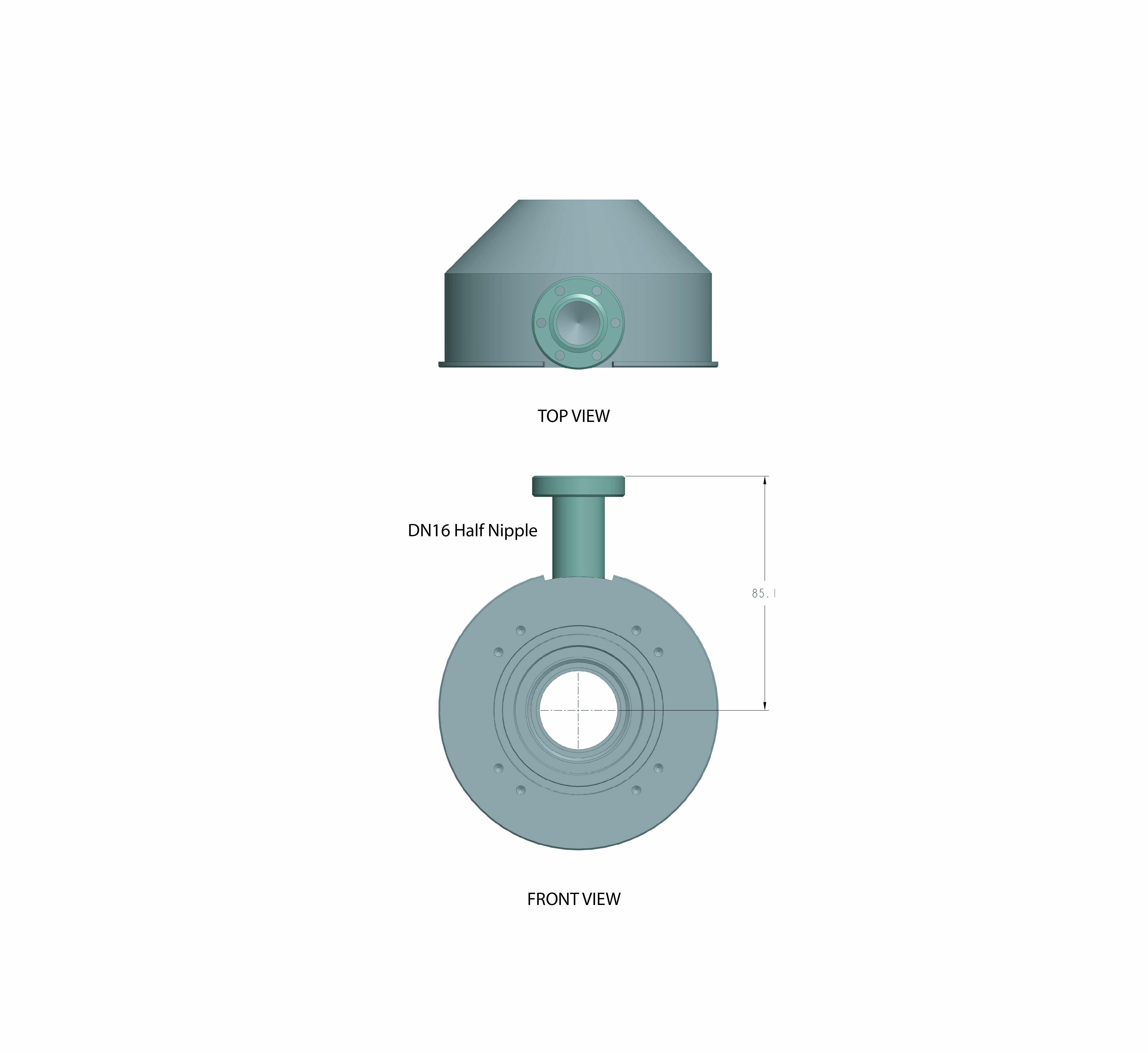}
     \caption{\textbf{Sleeve weldment.} Top and Front views of the sleeve after welding an appropriate DN16 conflat flange used for evacuating the secondary (sample to stage) volume between sample transfers.}
    \label{fig2:APXPS}
\end{figure}
